\pdfoutput=1

\documentclass[prl,twocolumn,twoside,preprintnumbers,superscriptaddress]{revtex4}

\usepackage{amsmath}
\usepackage{graphicx}
\usepackage[hyperfootnotes=false]{hyperref}
\usepackage{xspace}

\newcommand{\eq}[1]{Eq.~\eqref{eq:#1}}

\newcommand{\fig}[1]{Fig.~\ref{fig:#1}}

\newcommand{\pythia}{{\sc Pythia}\xspace}
\newcommand{\fastjet}{\textsc{FastJet}\xspace}

\newcommand{\ord}[1]{\mathcal{O}(#1)}

\newcommand{\df}{\mathrm{d}}

\newcommand{\tr}{\mathrm{tr}}
\newcommand{\nn}{\nonumber}
\newcommand{\cJ}{\mathcal{J}}

\newcommand{\al}{\alpha}

\newcommand{\de}{\delta}
\newcommand{\eps}{\epsilon}
\newcommand{\si}{\sigma}

\newcommand{\zero}{{(0)}}
\newcommand{\one}{{(1)}}

\allowdisplaybreaks[4]

\begin{document}


\preprint{\vbox{\hbox{MIT--CTP 4449}}}

\title{Calculating Track-Based Observables for the LHC}

\author{Hsi-Ming Chang}

\affiliation{Department of Physics, University of California at San Diego, 
La Jolla, CA 92093, USA}

\author{Massimiliano Procura}

\affiliation{Albert Einstein Center for Fundamental Physics, Institute for Theoretical Physics, University of Bern, CH-3012 Bern, Switzerland}

\author{Jesse Thaler}

\affiliation{Center for Theoretical Physics, Massachusetts Institute of Technology, Cambridge, MA 02139, USA}

\author{Wouter J.~Waalewijn}

\affiliation{Department of Physics, University of California at San Diego, 
La Jolla, CA 92093, USA}

\begin{abstract}

By using observables that only depend on charged particles (tracks), one can efficiently suppress pile-up contamination at the LHC.  Such measurements are not infrared safe in perturbation theory, so any calculation of track-based observables must account for hadronization effects.  We develop a formalism to perform these calculations in QCD, by matching partonic cross sections onto new non-perturbative objects called track functions which absorb infrared divergences.  The track function $T_i(x)$ describes the energy fraction $x$ of a hard parton $i$ which is converted into charged hadrons.  We give a field-theoretic definition of the track function and derive its renormalization group evolution, which is in excellent agreement with the \pythia parton shower.   We then perform a next-to-leading order calculation of the total energy fraction of charged particles in $e^+ e^- \to$ hadrons.   To demonstrate the implications of our framework for the LHC, we match the \pythia parton shower onto a set of track functions to describe the track mass distribution in Higgs plus one jet events.   We also show how to reduce smearing due to hadronization fluctuations by measuring dimensionless track-based ratios.

\end{abstract}

\maketitle

Jets are collimated sprays of particles that arise from the fragmentation of energetic quarks and gluons.  Nearly every measurement at the Large Hadron Collider (LHC) involves jets in some way, either directly as probes of physics in and beyond the Standard Model, or indirectly as a source of backgrounds and systematic uncertainties.  In order to predict jet-based observables using quantum chromodynamics (QCD), one typically performs infrared- and collinear-safe (IRC safe) jet measurements which involve only the kinematics of the jet constituents~\cite{Salam:2009jx}.  In particular, IRC safe jet measurements do not distinguish between charged and neutral particles, despite the fact that, for example, charged pions ($\pi^\pm$) are measured using both tracking and calorimetry whereas neutral pions ($\pi^0$) are measured using calorimetry alone.

In this letter, we develop the theoretical formalism to calculate track-based observables, which depend on the kinematics of charged particles alone but not on their individual properties or multiplicities.  The experimental motivation for track-based measurements is that tracking detectors offer better pointing and angular resolution than calorimetry.  By only using tracks, one can substantially mitigate the effects of pileup (multiple collision events in a single bunch crossing) which is becoming more relevant as the LHC achieves higher luminosity (see e.g.~\cite{Cacciari:2007fd,Krohn:2009th,Ellis:2009me,Alon:2011xb,Soyez:2012hv} for alternative approaches).   In addition, tracks can aid in jet substructure studies where the angular energy distribution in the jet discriminates between different jet types \cite{Abdesselam:2010pt,Altheimer:2012mn}.  While we focus on charged particles, this formalism applies to any (otherwise) IRC safe measurement performed only on a subset of particles.

To describe track-based observables in QCD, we introduce the \emph{track function} $T_i(x,\mu)$. A parton (quark or gluon) labelled by $i$ with four-momentum $p_i^\mu$ hadronizes into charged particles with total four-momentum $\overline{p}_i^\mu \equiv x p_i^\mu+\ord{\Lambda_\text{QCD}}$. The distribution in the energy fraction $0 \le x \le 1$ is the track function and is by definition normalized
\begin{align} \label{eq:T_norm}
 \int_0^1\! \df x\ T_i(x,\mu) = 1
\,.\end{align}
The track function is similar to a fragmentation function (FF) or a parton distribution function (PDF) in that it is a fundamentally non-perturbative object that absorbs infrared (IR) divergences in partonic calculations.  Like FFs and PDFs, the track function has a well-defined dependence on the renormalization group (RG) scale $\mu$ through a DGLAP-type evolution~\cite{Gribov:1972ri, Georgi:1951sr, Gross:1974cs, Altarelli:1977zs, Dokshitzer:1977sg}, though the specific evolution is more reminiscent of the jet charge distribution \cite{Krohn:2012fg,Waalewijn:2012sv}. For the observables we consider, each parton has its own independent track function. Hadronization correlations are captured by power corrections (beyond the scope of this letter).

\begin{figure}
\includegraphics[height=25ex]{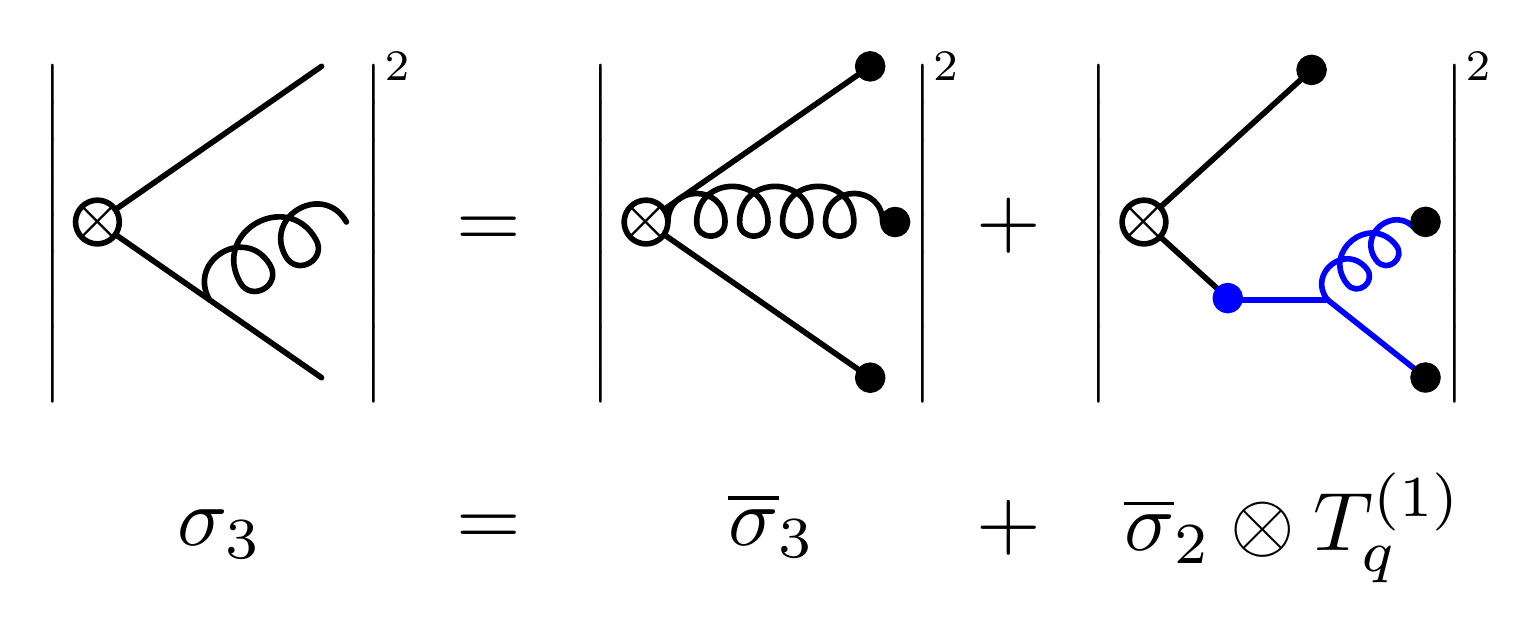} \\[-1ex]
\caption{Schematic relationship between the partonic matrix element $\sigma_3$ and the matching coefficient $\overline{\sigma}_3$ for $e^+ e^- \to q \bar{q} g$.   Here, black (blue) dots represents the tree-level ($\mathcal{O}(\alpha_s)$) track functions.  Diagrams with emissions from the other quark leg are elided for simplicity.  Note the trivial matching condition $\sigma_2 = \overline{\sigma}_2$. 
\vspace{-1ex}}
\label{fig:matching}
\end{figure}

Consider the cross section for an IRC safe observable $e$ measured using partons
\begin{align}
\frac{\df \sigma}{\df e} = \sum_N \int\! \df \Pi_N\, \frac{ \df \si_N}{\df \Pi_N}\, \de[e - \hat{e}(\{p^\mu_i\})]
\,,\end{align}
where we drop possible convolutions with PDFs to keep the notation simple.  Here, $\Pi_N$ denotes $N$-body phase space, $\df \si_N/\df \Pi_N$ is the corresponding partonic cross section, and $\hat{e}(\{p_i\})$ implements the measurement on the partonic four-momenta $p^\mu_i$.  Since $e$ is an IRC safe observable, the KLN theorem~\cite{Kinoshita:1962ur,Lee:1964is} guarantees a cancellation of final state IR divergences between real and virtual diagrams.  
The cross section for the same observable measured using only tracks is 
\begin{equation} \label{eq:trackxsec}
\frac{\df \sigma}{\df \overline{e}}= \sum_N \int\! \df \Pi_N\, \frac{\df \bar \si_N}{\df \Pi_N} \int\! \prod_{i=1}^N \df x_i\, T_i(x_i)\, \de[\bar e - \hat{e}(\{x_i p^\mu_i\})],
\end{equation}
where $T_i(x_i)$ is the track function for parton $i$. This equation defines a matching onto track functions where $\df \bar \si_N/\df \Pi_N$ represents the short distance matching coefficient, which is calculable in perturbation theory. In the absence of track functions, $\df \sigma/ {\df \overline{e}}$ would exhibit a mismatch between real and virtual diagrams in the form of uncompensated IR divergences in the partonic computation. The track functions absorb these IR divergences, and the partonic cross section $\bar \si_N$ is correspondingly modified with respect to $\si_N$. We will show below for the example of $e^+ e^- \to q \bar{q} g$ how the mismatch in the absence of $T_i(x_i)$ occurs. \fig{matching} shows schematically how we determine the IR-finite matching coefficient $\bar \si_3$ for this case, by using that \eq{trackxsec} is valid both at the hadronic and partonic level.
The fact that we consider factorizable (otherwise) IRC-safe observables modified to include only charged particles and that collinear divergences are known to be universal in QCD~\cite{Berends:1987me,Mangano:1990by,Kosower:1999xi} guarantees a valid matching to all orders in the strong coupling constant $\alpha_s$.

\begin{figure}
\includegraphics[height=35ex]{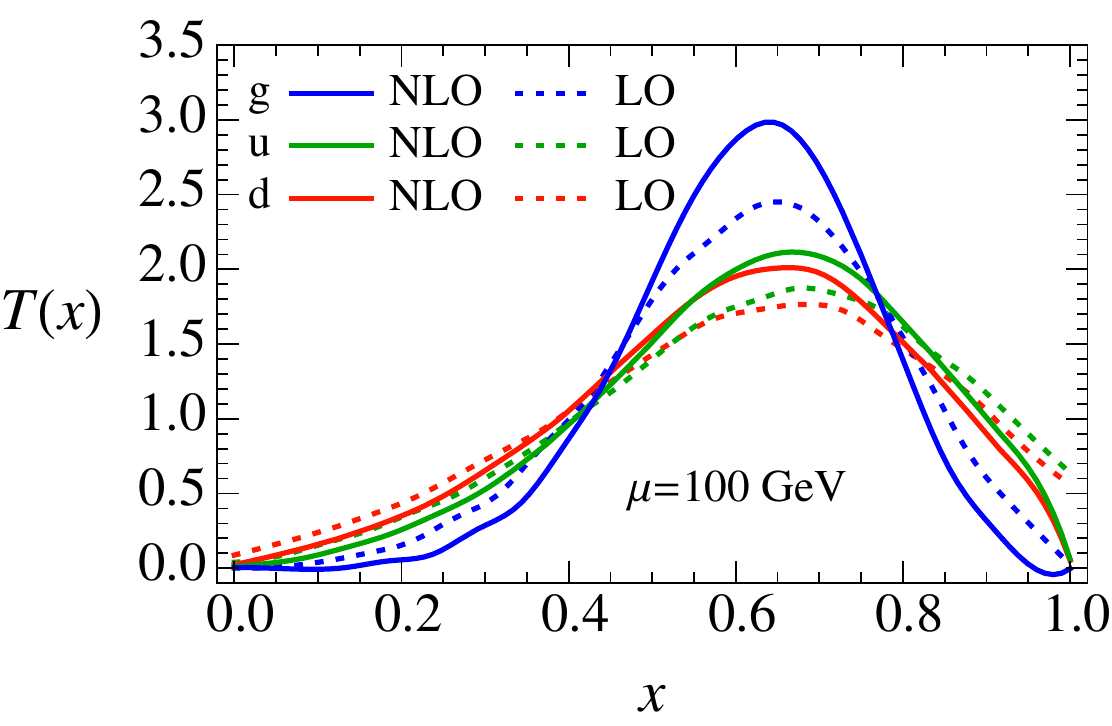} \\[-1ex]
\caption{LO (dotted) and NLO (solid) track functions extracted in \pythia from the fraction of the jet energy carried by charged particles. \vspace{-1ex}}
\label{fig:pythiatrack}
\end{figure}

At leading order (LO) in $\alpha_s$, the cross section depends on a single partonic multiplicity $N$ and there are no IR divergences implying $\bar{\sigma}^{(0)}_N = \sigma^{(0)}_N$. The LO $T_i^{(0)}(x_i)$ is simply a finite distribution which can be obtained directly from the energy fraction of charged particles in a jet initiated by a parton $i$. Ideally, we would extract this information from data, but just for illustrative purposes, we can determine it from (tuned) Monte Carlo event generators. We stress that our formalism does not rely on the use of these programs nor on their built-in hadronization models. In \fig{pythiatrack}, we show the track functions obtained from pure quark and gluon jet samples produced by \pythia 8.150~\cite{Sjostrand:2006za,Sjostrand:2007gs} and clustered using the anti-$k_T$ algorithm~\cite{Cacciari:2008gp} in \fastjet 2.4.4 \cite{Cacciari:2011ma}.  (To extract the track function at next-to-leading order (NLO) we use \eq{jet_NLO}; the jet radius $R$ is correlated with the RG scale $\mu$.)  As expected, the up- and down-quark track functions are very similar, with a peak at $x = 0.6$. This means that on average 60\% of the energy of the initial quark is contained in charged hadrons, in agreement with a recent CMS study~\cite{CMS:2010rua}.  The small difference between up and down is due to strangeness, since $u \bar{s}$ mesons are charged whereas $d \bar{s}$ mesons are neutral.  Because gluons have a larger color factor than quarks, they yield a higher track multiplicity, and the corresponding track functions are narrower, as expected from the central limit theorem.

Formally, the (bare) track function is defined in QCD in a fashion analogous to the unpolarized FF (cf.~\cite{Collins:1981uk,Collins:1981uw}).  Expressed in terms of  light-cone components,
\begin{align} \label{eq:defTqQCD} 
 T_{q}(x)&=
\int\! \df y^+\,\df^2 y_\perp\;e^{\,i k^- \, y^+/2}  \, \frac{1}{2 N_c}\, \sum_{C, N} \de\Big(x - \frac{p_C^-}{k^-}\Big)
\nn \\ & \, \, \times \tr\Big[ \frac{\gamma^-}{2}\,
  \langle 0 | \psi (y^+, 0, y_\perp )| C N
  \rangle \langle C N | \overline{\psi}(0) |0 \rangle \Big]
\,,\end{align}
where $\psi$ is the quark field, $C$ ($N$) denote charged (neutral) hadrons, and $p_C^-$ is the large momentum component of all charged particles.  Whereas the FF describes the energy fraction carried by an individual hadron, the track function describes the energy fraction carried by all charged particles.  As for the FF, gauge invariance requires the addition of eikonal Wilson lines.  The gluon track function is defined analogously~\cite{Chang:2013iba}.

\begin{figure}
\includegraphics[height=35ex]{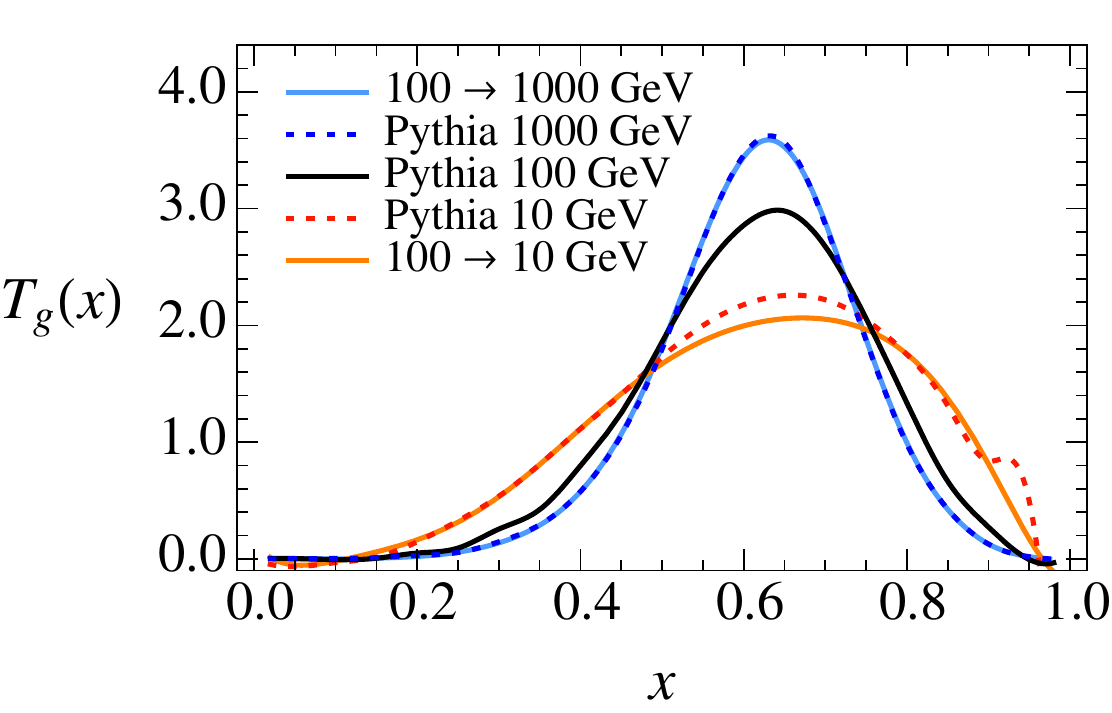} \\
\includegraphics[height=35ex]{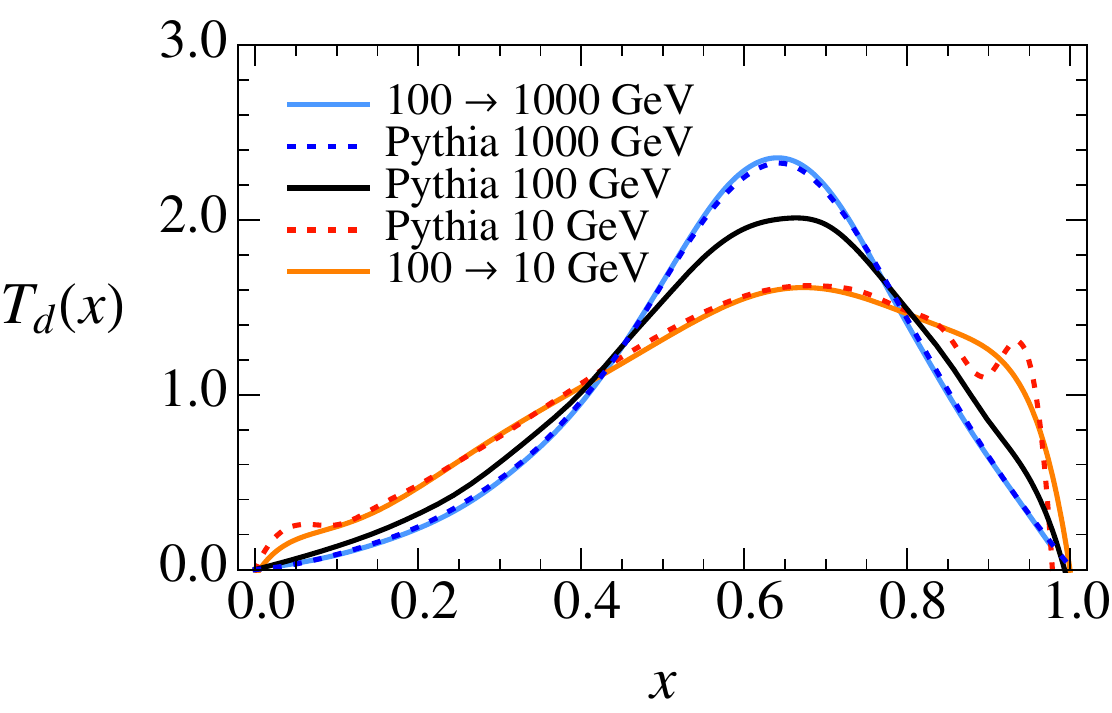} \\[-1ex]
\caption{The evolution of the NLO gluon (top) and $d$-quark (bottom) track functions compared to \pythia.  Starting from $\mu=100$ GeV (shown in \fig{pythiatrack}), we evolve using \eq{T_RGE} down to $\mu=10$ GeV and up to $\mu=1000$ GeV.  The bumps in the \pythia\ distributions near $x=0,1$ at $Q = 10$ GeV correspond to genuine non-perturbative effects at $\Lambda_{\rm QCD}$. \vspace{-1ex}}
\label{fig:evolution}
\end{figure}

Treating the intermediate states in \eq{defTqQCD} partonically, we obtain the bare track functions $T_{i,\text{bare}}^\one$ at NLO in pure dimensional regularization with $d = 4 - 2 \epsilon$,
\begin{align} \label{eq:Toneloop}
T_{i,\text{bare}}^\one(x) & = \frac{1}{2} \sum_{j,k} \int\! \df z \, \Big[\frac{\al_s(\mu)}{2\pi} \Big(\frac{1}{\eps_\text{UV}} - \frac{1}{\eps_\text{IR}} \Big) P_{i \to jk}(z)\Big] 
\nn \\ & \quad \times 
\int \! \df x_1 \, \df x_2 \,  T^\zero_j(x_1,\mu) T^\zero_k(x_2,\mu) 
\nn \\ & \quad \times 
\de\big[x - z x_1 - (1-z) x_2\big]
\,,\end{align}
which arise from collinear splittings, controlled by the timelike Altarelli-Parisi splitting functions $P_{i \to jk}(x)$~\cite{Altarelli:1977zs}. In contrast with the analogous partonic FF calculation, track functions involve contributions from both branches of the splitting. Renormalizing the ultraviolet divergences in \eq{Toneloop} in $\overline{\text{MS}}$ leads to the evolution equation for the track function
\begin{align}\label{eq:T_RGE}
  \mu \frac{\df}{\df \mu}\, T_i(x, \mu) &= \frac{1}{2} \sum_{j,k} \int\! \df z\, \df x_1\, \df x_2\, \frac{\al_s(\mu)}{\pi} P_{i \to j k}(z)
  \\ & \quad \times 
  T_j(x_1,\mu) T_k(x_2,\mu)\,\de[x \!-\! z x_1 \!-\! (1\!-\!z) x_2]. 
\nn\end{align}
Like for a PDF, the track function can be extracted at one scale and RG evolved to another scale, and the evolution preserves the normalization in \eq{T_norm}.  Unlike a PDF, \eq{T_RGE} involves a convolution of two track functions at NLO (and more convolutions at higher orders corresponding to multiple branchings), so it is numerically more involved to perform the $\mu$-evolution.   At leading logarithmic (LL) order, the RG evolution in \eq{T_RGE} is equivalent to a parton shower, and \fig{evolution} demonstrates excellent agreement between our numerical evolution and the parton shower in \pythia. 

\begin{figure}[t]
\includegraphics[height=35ex]{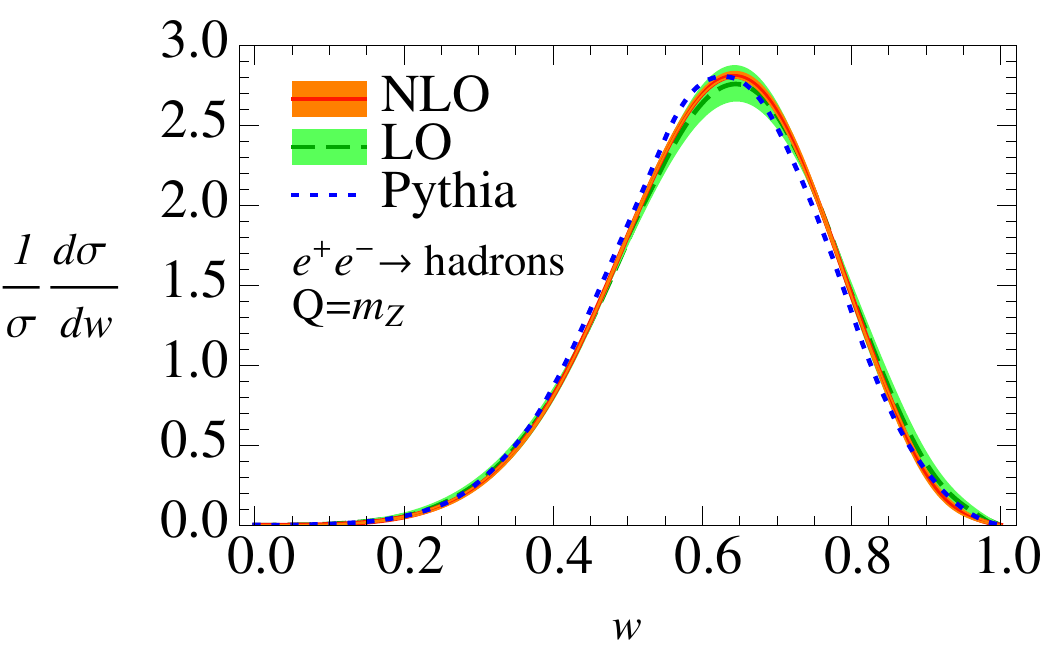} \\[-1ex]
\caption{Normalized distribution of the energy fraction $w$ of charged particles in $e^+ e^-$ at $Q=91$ GeV, calculated at LO (green) and NLO (orange), compared with \pythia (blue). The uncertainty bands are obtained by varying $\mu$ between $Q/2$ and $2 Q$, and do not include track function uncertainties.\vspace{-1ex}}
\label{fig:epemenergyfraction}
\end{figure}

For a calculation at NLO, both the partonic cross section and the track functions have IR divergences which cancel in \eq{trackxsec}. To demonstrate this in a simple example, consider the process $e^+ e^- \to \text{hadrons}$ at a center-of-mass energy $Q$ where one measures the total energy fraction $w$ of charged particles.  At NLO the partonic process is $e^+ e^- \to q \bar{q} g$, whose kinematics are described by the energy fractions $y_1$ and $y_2$ of the quark and anti-quark.  Applying \eq{trackxsec} we find
\begin{align}
\label{eq:epemwxsec}
  \!\!\frac{\df \si}{\df w} &= \int\! \df y_1 \df y_2\, \frac{\df \bar \si}{\df y_1 \df y_2} 
  \int \! \df x_1 \df x_2  \df x_3 \,T_q(x_1) T_q(x_2) T_g(x_3)
  \nn \\  &\quad 
  \times \de(w - [y_1 x_1 + y_2 x_2 + (2-y_1-y_2) x_3]/2)
,\end{align}
since $T_q = T_{\bar{q}}$. The matching coefficient $\df \bar \si$ is extracted by evaluating this equation at the partonic level (see  \fig{matching}). We then use it in \eq{epemwxsec} together with non-perturbative hadronic track functions to obtain the physical cross section $\df \si / \df w$.   At the LO partonic level
\begin{align}
\frac{\df \si^\zero}{\df y_1 \df y_2} &= \si^\zero\, \de(1-y_1) \de(1-y_2)
,
\end{align}
where $\si^\zero$ is the total Born cross section. At NLO, the cross section can be expressed using plus-functions as
\begin{widetext}
\ \\[-6ex]
\begin{align} 
\label{eq:ee_si}
  \frac{\df \si^\one}{\df y_1 \df y_2}
  &= \si^\zero\, \frac{\al_s(\mu) C_F}{2\pi} \bigg\{\Big(\frac{\pi^2}{2}-4\Big) \de(1-y_1) \de(1-y_2) 
   + \frac{\theta(y_1+y_2-1)(y_1^2 + y_2^2)}{2(1-y_1)_+(1-y_2)_+}
   \\ & \quad
  + \de(1-y_1) \Big[-\frac{1}{\eps_\text{IR}} \frac{P_{q\to qg}(y_2)}{C_F}  
  + (1+y_2^2) \Big[\frac{\ln(1-y_2)}{1-y_2}\Big]_+ \!\!+ \frac{P_{q\to qg}(y_2)}{C_F} \ln \frac{y_2 \, Q^2}{\mu^2}  + 1 - y_2\Big] 
  + (y_1 \leftrightarrow y_2)  \bigg\}
\,, \nn
\end{align}
\end{widetext}
where $C_F=4/3$ and both real and virtual contributions are included.
The $1/\epsilon_\text{IR}$-divergences in $\si^\one$ are cancelled by the ones in $T^\one$ from \eq{Toneloop}, and the finite remainder defines $\bar \si^\one$ in $\overline{\text{MS}}$. (For different IR regulators, the track function may also contribute finite terms to the matching.)   By performing this kind of matching calculation, one can determine $\bar \si_N$ for any process to any order in $\alpha_s$. For the case of \eq{ee_si} which involves a single scale $Q$, we choose $\mu \simeq Q$ to minimize the logarithms in $\bar \si$. In \fig{epemenergyfraction} we show the LO and NLO distributions for the energy fraction $w$, using the track functions extracted from \pythia by means of \eq{jet_NLO}. There is good convergence from LO to NLO and our fixed-order calculation agrees well with \pythia.

Ultimately, we are interested in applying the track function formalism to jet-based measurements at the LHC, which typically involve multiple scales. For narrow well-separated jets, though, the contributions from soft radiation are power-suppressed, so the energy fraction $x$ of the charged particles within a single jet depends on the jet scale
\begin{equation} \label{eq:mu_J}
\mu_J \simeq p_T R
\,,\end{equation}
where $p_T$ is the transverse momentum of the jet and $R$ denotes its azimuthal-rapidity size as defined by a jet algorithm (anti-$k_T$ in this letter).  Indeed, by varying $R$ (and trying different jet algorithms) while keeping $\mu_J = p_T R$ fixed, we find nearly identical $x$ distributions in \pythia.   At LO, the normalized distribution in $x$ defines $T_i^{(0)}(x)$ itself, shown in \fig{pythiatrack}. At NLO, the distribution of the energy fraction $x$ within a jet initiated by parton $i$ has the same form as for the jet charge distribution~\cite{Waalewijn:2012sv}
\begin{align} \label{eq:jet_NLO}
\frac{1}{\si_i} \frac{\df \si_i}{\df x} &= \frac{1}{2} \sum_{j,k} \int\! \df x_1\, \df x_2\, \df z\, \frac{\cJ_{ij}(p_T R,z,\mu_J)}{2(2\pi)^3\,J_i(p_T R,\mu_J)} 
 \\ & \quad \times
 T_j(x_1,\mu_J) T_k(x_2,\mu_J) \, \de[x\!-\!z x_1\!-\!(1-z) x_2]
\,. \nn \end{align}
The jet functions $J_i(p_T R,\mu)$~\cite{Ellis:2010rwa} arise from the $1/\si_i$ normalization factor and describe jet production without any additional measurement. The matching coefficients are $\cJ_{ij}$~\cite{Procura:2011aq,Waalewijn:2012sv}. At NLO there is at most a $1\to2$ splitting which completely fixes the kinematics, so the same $\cJ_{ij}$ appear in the jet charge~\cite{Waalewijn:2012sv} and in fragmentation inside an identified jet~\cite{Procura:2009vm,Jain:2011xz}. The coefficients will not be the same at higher orders. We can invert \eq{jet_NLO} to determine the NLO track functions in \fig{pythiatrack}.

As a track-based measurement relevant for the LHC, consider the track-only jet mass spectrum in $pp$ to Higgs plus one jet.  This example is more complicated than the $e^+e^-$ example above since it involves several scales:  the hard scale set by the transverse momentum of the jet $p_T^J$, the jet mass scale $m_J$, and the scale associated with soft radiation $m_J^2/p_T^J$.  In a resummed jet mass calculation, one would need to treat each emission in the exponentiated soft function as hadronizing independently, i.e.~convolved with its own track function~\cite{Chang:2013iba}.  At LL order, however, we can use the fact that the parton shower already describes the numerous parton emissions that build up the jet mass distribution.  Thus, a correct and instructive use of our formalism at LL is to run the perturbative \pythia parton shower down to a low scale $\mu \simeq \Lambda_{\rm QCD}$ and match each final state parton onto a track function (also evolved to $\mu \simeq \Lambda_{\rm QCD}$).  

\begin{figure}
\includegraphics[height=35ex]{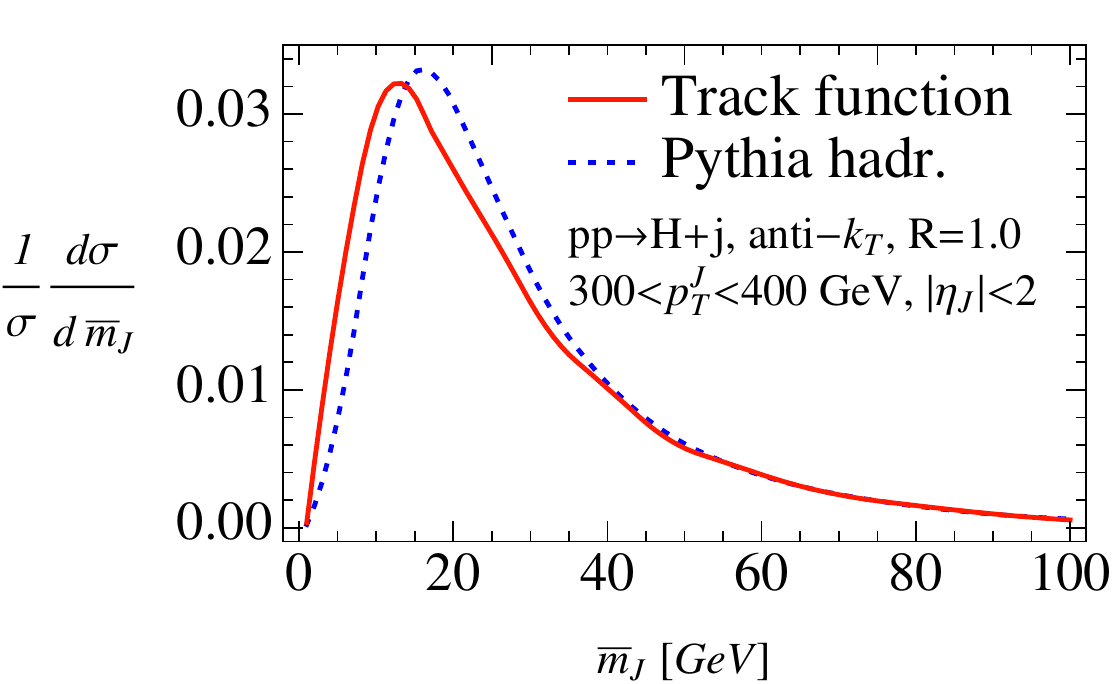} \\[-1ex]
\caption{Track mass distribution in $pp \to H+$jet obtained from the \pythia parton shower matched onto either track functions or the Lund string model. \vspace{-1ex}}
\label{fig:LHCtrackmass}
\end{figure}

In \fig{LHCtrackmass}, we show the track mass $\bar m_J$ spectrum using the E-scheme~\cite{Salam:2001bd} for the anti-$k_T$ jet algorithm with $R=1.0$. We impose realistic cuts on the jet rapidity $\eta_J$ and $p_T^J$ at the calorimeter level (using all the particles), to select the same events for both curves. There is excellent agreement between the track function and the Lund hadronization model \cite{Andersson:1983ia} in \pythia, which is a non-trivial check considering that the track functions were originally extracted from \pythia using jet energies (and not jet mass) at $\mu = 100$ GeV. Since the track function does not include correlations between the hadronization of different partons, the agreement in \fig{LHCtrackmass} shows that independent fragmentation is an excellent approximation for these observables. The small difference between the two distributions in the peak region is due to nonperturbative power corrections, which to first approximation can be described by a shift in $m_J^2$ (see e.g.~\cite{Dokshitzer:1997ew,Salam:2001bd,Lee:2006nr,Mateu:2012nk}).

\begin{figure}
\includegraphics[height=35ex]{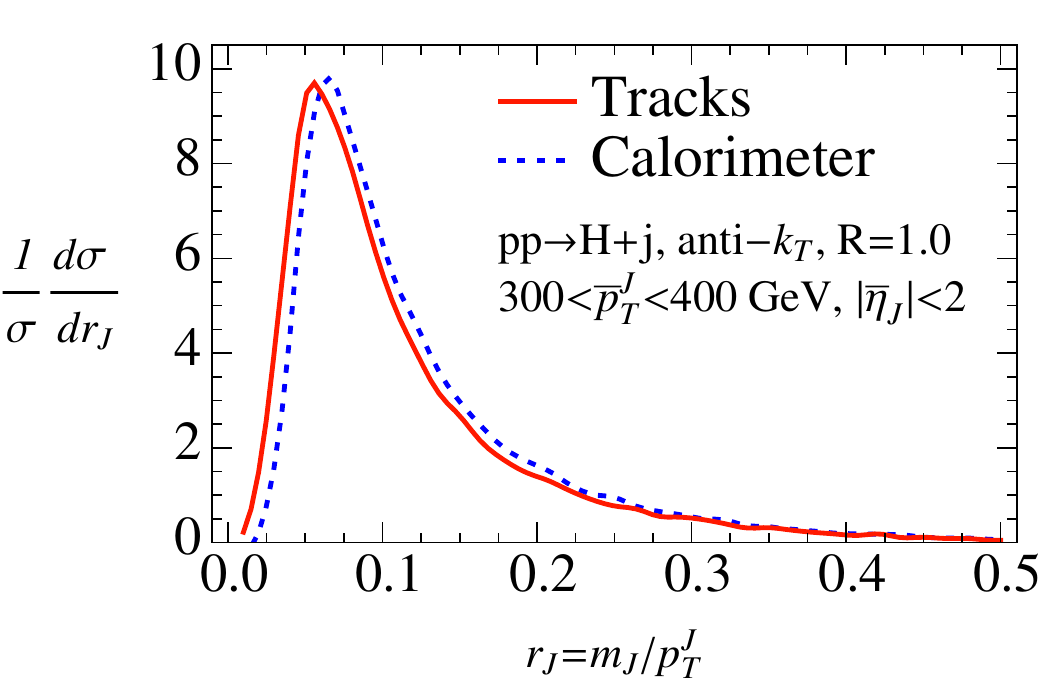} \\[-1ex]
\caption{Comparison between track and calorimeter measurements of ratio of jet mass and jet $p_T$.
Here the cuts are on $\bar p_T^J$ and $\bar \eta_J$ from the tracks in the jet. \vspace{-1ex}}
\label{fig:LHCratio}
\end{figure}

Despite the advantages of using track-based observables for pile-up suppression, it should be noted that the track functions have a rather large width (see \fig{pythiatrack}).  
Fluctuations in the charged energy fraction produce an effective energy smearing for a track-based measurement relative to a calorimetric one.  To partially address this issue, one can focus on observables for which the track function would only have a modest effect, such as for dimensionless ratios of observables.  A particularly useful example are $N$-subjettiness ratios \cite{Thaler:2010tr}, which are relevant for jet substructure studies.  Though a calculation of such ratios is beyond the scope of this letter, we plot in \fig{LHCratio} the ratio of the jet mass to jet $p_T$ in \pythia, measured using either tracks alone or all particles (calorimeter).  As expected, the smearing effect is reduced since hadronization fluctuations are correlated between the numerator and denominator. 

Given the potential experimental benefits from track-only measurements at high luminosities, we expect that track functions will offer an important theoretical handle for future precision jet studies at the LHC. Beyond the tests of our formalism against \pythia performed here, we stress that it is possible to systematically improve the accuracy of track-based predictions.  At fixed order in $\alpha_s$, one can calculate higher-order corrections for the (process-dependent) matching onto track functions.  More ambitiously, one could match track functions onto automated NLO calculations, perhaps using a more convenient IR regulator (such as dipole subtraction \cite{Catani:1996vz}) rather than the $\overline{\text{MS}}$ scheme used here.  To improve the accuracy of resummation, one needs to determine higher-order RG evolution of the track functions, and precision track-based studies would also require power corrections.   Ultimately, one would want to follow the example of PDFs and extract track functions for the LHC from global fits to data.

\begin{acknowledgments}
We thank A.~Manohar and I.~Stewart for discussions and feedback on this manuscript. We also thank M.~Pierini for discussions. H.C.~and W.W.~are supported by DOE grant DE-FG02-90ER40546. M.P.~acknowledges support  by the ``Innovations- und Kooperationsprojekt C-13'' of the Schweizerische Universit\"atskonferenz SUK/CRUS and by the Swiss National Science Foundation. J.T.~is supported by DOE grants DE-FG02-05ER-41360 and DE-FG02-11ER-41741.
\end{acknowledgments}

\bibliography{tracks}

\end{document}